# Correlations, inhomogeneous screening, and suppression of spin-splitting in quantum wires at strong magnetic fields

O. G. Balev[*] and P. Vasilopoulos[†]

[*]Institute of Physics of Semiconductors, National Academy of Sciences, 45 Pr. Nauky, Kiev 252650, Ukraine

[†]Concordia University, Department of Physics, 1455 de Maisonneuve Blvd O, Montréal, Québec, Canada, H3G 1M8

(September 26, 1996)

## Abstract

A self-consistent treatment of exchange and correlation interactions in a quantum wire (QW) subject to a strong perpendicular magnetic field is presented using a modified local-density approximation (MLDA). The influence of many-body interactions on the spin-splitting between the two lowest Landau levels (LLs) is calculated within the screened Hartree-Fock approximation (SHFA), for filling factor $\nu = 1$, and the strong spatial dependence of the screening properties of electrons is taken into account. In comparison with the Hartree-Fock result, the spatial behavior of the occupied LL in a QW is strongly modified when correlations are included. Correlations caused by screening at the edges strongly suppress the exchange splitting and smoothen the energy dispersion at the edges. The theory accounts well for the experimentally observed strong suppression of the spin-splitting pertinent to the $\nu = 1$ quantum Hall effect (QHE) state as well as the destruction of this state in long, quasi-ballistic GaAlAs/GaAs QWs.

# I. INTRODUCTION.

Recently the effects of electron-electron interactions on the edge state properties of a channel [1]–[5] and on the subband structure of QWs [6]–[8], [4] have attracted significant attention. One consensus of the theoretical work is that it is important to include the Coulomb interactions self-consistently. In the present work we introduce a realistic model of a QW in a strong magnetic field $B$ and self-consistently treat mainly the case when the lowest, spin-polarized, LL is occupied, i.e., when $\nu = 1$ in the interior part of a channel and, in the assumed integral QHE regime, the formation of a dipolar strip [1] at the channel edges is impossible. Moreover, we consider submicron width channels with rather steep confining potential that prevents the flattening of edge states [2], [4], [7] in the vicinity of the Fermi level that was suggested in Ref. [1]. To date we are aware of only the Hartree [7], [4] and Hartree-Fock [6] treatments of LLs in a QW, at a strong $B$ field, that are similar to the edge-state studies of a wide channel [1]–[4]. Here we show that, if we include correlations in the Coulomb interaction in a QW, the spatial behavior of the LLs is strongly modified.

We use the SHFA [9] to take into account exchange and correlation effects in calculating the LL single-particle energies and assessing the spatial dependence of the spin-splitting. Including correlations leads to strong changes in the spin-splitting between the two lowest LLs. These changes differ essentially between the middle of the channel and the region near the edges. The most essential role played by correlations is related with screening by the edge states which in turn depends strongly on their (group) velocity $v_g$. The correlations can restore a smooth, on the scale of the magnetic length $\ell_0 = (\hbar/m^*\omega_c)^{1/2}$, dispersion of the single-particle energy as a function of the oscillator center $y_0 \approx k_x \ell_0^2$ where $\omega_c$ is the cyclotron frequency. It is assumed that the confining potential without many-body interactions is smooth on the $\ell_0$ scale and, hence, leads to a rather small $v_g^H$; notice that in this case the exchange interaction leads to an infinite (logarithmically divergent) $v_g$. Because in typical experimental situations the *strong magnetic field limit* condition, $r_0 = e^2/(\varepsilon \ell_0 \hbar \omega_c) \ll 1$, is not satisfied, we propose a modified local-density approximation (MLDA) to self-consistently treat the effect of many-body interactions in a *strong B*, when $r_0 \lesssim 1$. Without correlation effects and in a *strong magnetic field limit* our model is similar to that of Ref. [6].

For integer $\nu$ the self-consistent confining potential is close to being parabolic [7] . In addition, in Ref. [7] it was shown that the overall distribution of electron charges is primarily determined by the large electrostatic energy and remains almost independent of the field B although the confining potential and the subband dispersion can change drastically with $B$. Because of this here we do not treat the suppression of spin-splitting due to effect considered in Ref. [6] that requires strong changes in the total distribution of electron charges and, hence, in the electrostatic (Hartree) energy. We typically assume $\omega_c/\Omega \gg 1$, where $\Omega$ is the confining frequency. This implies a sufficiently strong field $B$ whereas the effect considered in Ref. [6] normally requires $\omega_c \sim \Omega$. Our theory describes well the experimentally observed spin-splitting in GaAlAs/GaAs QWs [8].

In Sec. II we present the basic formalism and in a *strong magnetic field limit* show how the single-particle energies are modified when many-body interactions are included. In Sec. III we show how strong correlations result from screening at the edges. In Sec. IV we propose, for strong $B$, a MLDA and obtain, within its framework, a strong suppression of the exchange splitting, the restoration of a smooth energy dispersion at the edges, and the possibility of destruction of the $\nu = 1$ QHE state. In addition, in Sec. IV we apply our theory to the experimental results of Ref. [8]. We conclude with remarks in Sec. V.

## II. BASIC RELATIONS

### A. Channel characteristics without many-body interactions

We consider a two-dimensional electron gas (2DEG) confined in a narrow channel, in the $(x, y)$ plane, of width $W$ and of length $L_x = L$; for simplicity we will neglect its thickness $d$. In the absence of exchange and correlation effects we take the confining potential along $y$ as parabolic: $V_y = m^*\Omega^2 y^2/2$, where $m^*$ is the effective mass. When a strong field $B$ is applied along the z axis, in the Landau gauge for the vector potential $\mathbf{A} = (-By, 0, 0)$ the one-electron Hamiltonian $\hat{h}^0$ is given by

$$\hat{h}^0 = [(\hat{p}_x + eBy/c)^2 + \hat{p}_y^2]/2m^* + V_y + g_0\mu_B \hat{S}_z B/2, \qquad (1)$$

where $\hat{\mathbf{p}}$ is the momentum operator, $e(<0)$ the electron charge, $g_0$ the bare Landé g-factor, and $\mu_B$ the Bohr magneton. $\hat{S}_z$ is the z-component of the spin operator with eigenvalues $\sigma = 1$ and $\sigma = -1$ for spin $\uparrow$ and $\downarrow$, respectively. As is usual, we consider a parabolic lateral confinement, used especially for $W \lesssim 0.3\mu$m, cf. Refs. [6] and [8]. However, as will be indicated below most of our results hold for the potential $V_y' = 0$, for $y_l < y < y_r$, $V_y' = m^*\Omega^2(y - y_r)^2/2$ for $y > y_r > 0$, and $V_y' = m^*\Omega^2(y - y_l)^2/2$ for $y < y_l < 0$. $V_y'$ is a more realistic approximation to the confining potential, which is the sum of the bare confining potential and the Hartree potential, when the Fermi level, within the interior part of a channel, lies within the top occupied LL, cf. Ref. [7].

The eigenvalues and eigenfunctions corresponding to Eq. (1) are given, respectively, by

$$\epsilon_\alpha \equiv \epsilon_{n,k_x,\sigma} = \hbar\tilde{\omega}(n + 1/2) + \frac{\hbar^2 k_x^2}{2\tilde{m}} + \sigma g_0\mu_B B/2, \qquad (2)$$

and

$$|\alpha> \equiv |nk_x>|\sigma> = e^{ik_x x}\Psi_n(y - y_0(k_x))|\sigma>/\sqrt{L}. \qquad (3)$$

Here, $\tilde{\omega} = (\omega_c^2 + \Omega^2)^{1/2}$, $\omega_c = |e|B/m^*c$, $\tilde{m} = m^*\tilde{\omega}^2/\Omega^2$, $y_0(k_x) = \hbar\omega_c k_x/m^*\tilde{\omega}^2$, and $\Psi_n(y)$ is a harmonic oscillator function; $|\sigma> = \psi_\sigma(\sigma_1) = \delta_{\sigma\sigma_1}$ is the spin-wave function and $\sigma_1 = \pm 1$. For the calculations that will follow we need ($\vec{q} = \{q_x, q_y\}$) the matrix elements

$$<n'k_x'|e^{i\vec{q}\cdot\vec{r}}|nk_x> = (n'k_x'|e^{iq_y y}|nk_x)\delta_{q_x,-k_-} = (\frac{n'!}{n!})^{1/2}(\frac{aq_x + iq_y}{\sqrt{2}/\tilde{\ell}})^m e^{-u/2}L_{n'}^m(u)e^{iaq_y k_+ \tilde{\ell}^2/2}\delta_{q_x,-k_-}. \qquad (4)$$

Here $k_\pm = k_x \pm k_x'$, $m = n - n'$, $a = \omega_c/\tilde{\omega}$, $u = [a^2 q_x^2 + q_y^2]\tilde{\ell}^2/2$, $\tilde{\ell} = (\hbar/m^*\tilde{\omega})^{1/2}$ is the renormalized magnetic length, and $L_{n'}^m(u)$ the Laguerre polynomial. In agreement with experiments [8] we assume a sufficiently smooth lateral confinement such that $\Omega \ll \omega_c$, i.e., the confining potential affects the eigenfunctions $|\alpha>$ very little, but it substantially changes the eigenvalues. This condition is usually fulfilled if $B$ is not too weak [10]. Then $\tilde{\ell}$ practically coincides with the bare magnetic length $\ell_0 = (\hbar/m^*\omega_c)^{1/2}$. The matrix elements $|<n'k'_x|e^{i\vec{q}\cdot\vec{r}}|nk_x>|^2$ given by Eq. (4) coincide with those of Ref. [10]; they also coincide with those given by Eq. (7) of Ref. [6] if we assume $a \equiv 1$.

## B. Exchange and correlations in a narrow channel

In the strong magnetic field limit, see, e.g., Refs. [2], [3], and [6], the condition $r_0 = e^2/\varepsilon\ell_0\hbar\omega_c \ll 1$ should be satisfied [11], where $\varepsilon$ is the background dielectric constant. However, it is known that in the most important experiments the parameter $r_0$ is of order unity [11], cf. Refs. [5]- [8]. Nevertheless, it is believed that calculations in this limit provide a useful framework also when $r_0$ is of order unity, see, e.g., [2] and [6]. Further, if not specified, we assume that only the lowest spin level $\uparrow$ of the $n = 0$ LL is occupied.

Without many-body interactions, the one-electron density matrix $\hat{\rho}^{(0)}$ is assumed diagonal, i.e., $<\alpha|\hat{\rho}^{(0)}|\beta> = f_\alpha \delta_{\alpha\beta}$, where $f_\alpha = 1/[1 + exp((\epsilon_\alpha - E_F)/k_B T)]$ is the Fermi-Dirac function and $E_F$ the Fermi level. We will assume $T = 0$. Now in all QWs considered below, when correlations are neglected, the highest occupied states of the $(n = 0, \sigma = 1)$ LL are below the bottom of the empty $(n = 0, \sigma = -1)$ LL. Then $E_F$ it is really the Fermi level common to both LLs. However, in the SHFA, as well as in the HFA, in general we can formally consider that the occupied $(n = 0, \sigma = 1)$ LL and the empty $(n = 0, \sigma = -1)$ LL have different quasi-Fermi levels. Of course such a state is not thermodynamically stable, see also below. Now, when correlations are taken into account, the $(n = 0, \sigma = -1)$ LL can be empty, in some of the QWs considered below, only if it has a quasi-Fermi level different than that of the $(n = 0, \sigma = 1)$ LL. Considering the exchange contribution in the SHFA and to first order in $r_0$, we obtain the exchange and correlation contributions to the single-particle energy $E_{0,k_x,1}$ as

$$\epsilon^{ec}_{0,k_x,1} = -\frac{1}{8\pi^3}\int_{-k_F}^{k_F} dk'_x \int_{-\infty}^{\infty} dq_y \int_{-\infty}^{\infty} dq'_y V^s(k_-, q_y; q'_y)(0k_x|e^{iq_y y}|0k'_x)(0k'_x|e^{iq'_y y}|0k_x), \quad (5)$$

where $E_{0,k_x,1} = \epsilon_{0,k_x,1} + \epsilon^{ec}_{0,k_x,1}$ and $k_F = (\tilde{\omega}/\hbar\Omega)[2m^*\Delta E_{F\uparrow}]^{1/2}$ is the characteristic wave vector such that this level is filled only for $|k_x| \leq k_F$; $\Delta E_{F\uparrow} = E_F - \hbar\tilde{\omega}/2 - g_0\mu_B B/2$. Here $V^s(q_x, q_y; q'_y)$ is the Fourier transform of the screened Coulomb interaction $V^s(x-x', y; y')$ between two electrons at points $(x, y)$ and $(x', y')$. Due to translational invariance in the $x$ direction we have

$$V^s(x-x', y; y') = \frac{1}{8\pi^3} \int_{-\infty}^{\infty} dq_x \int_{-\infty}^{\infty} dq_y \int_{-\infty}^{\infty} dq'_y e^{iq_x(x-x')+iq_y y+iq'_y y'} V^s(q_x, q_y; q'_y). \quad (6)$$

If we neglect the screening of the Coulomb interaction between two electrons, at $(x,y)$ and $(x',y')$, by all other electrons in the QW, then in Eq. (6) $V^s(x-x', y; y')$ should be changed to $V(|\vec{r}-\vec{r}'|) = e^2/\varepsilon|\vec{r}-\vec{r}'|$, $\vec{r} = \{x, y\}$, and $V^s(q_x, q_y; q'_y)$, in Eq. (5), to

$$V_0^s(q_x, q_y; q'_y) = \frac{4\pi^2 e^2}{\varepsilon q}\delta(q_y + q'_y), \quad (7)$$

where $q = \sqrt{q_x^2 + q_y^2}$. Using Eqs. (4) and (7) the right-hand side (RHS) of Eq. (5) gives the exchange contribution to the single-particle energy in the form

$$\epsilon^{ex}_{0,k_x,1} = -\frac{e^2}{2\pi\varepsilon} \int_{-k_F}^{k_F} dk'_x \int_{-\infty}^{\infty} dq_y \frac{e^{-(a^2 k_-^2 + q_y^2)\tilde{\ell}^2/2}}{\sqrt{k_-^2 + q_y^2}}, \quad (8)$$

which coincides with Eq. (10) of Ref. [6] for $a \equiv 1$. After intergration over $q_y$ in Eq. (8) we obtain

$$\epsilon^{ex}_{0,k_x,1} = -\frac{e^2}{2\pi\varepsilon\tilde{\ell}} \int_{\tilde{k}_x-\tilde{k}_F}^{\tilde{k}_x+\tilde{k}_F} dt\, e^{-(2a^2-1)t^2/4} K_0(t^2/4), \quad (9)$$

where $\tilde{k}_{x,F} = k_{x,F}\tilde{\ell}$ and $K_0(x)$ is a modified Bessel function. Eq. (9) is similar to Eq. (11) of Ref. [3], apart from a factor $1/2$, when $\omega_c^2 \gg \Omega^2$. In this case $a^2$ in Eq. (9) can be well approximated by 1 and the confining potential is smooth on the scale of $\ell_0$. From Eq. (9) for $\tilde{k}_F \gg 1$ and $\tilde{k}_F - |\tilde{k}_x| \gg 1$, we obtain

$$\epsilon^{ex}_{0,k_x,1} \approx \epsilon^{ex}_0 = -\sqrt{\frac{\pi}{2}} \frac{e^2}{\varepsilon\tilde{\ell}} R(\Omega/\omega_c), \quad (10)$$

and, for $\tilde{k}_x = \pm\tilde{k}_F$, $\epsilon^{ex}_{0,\pm k_F,1} = \epsilon^{ex}_0/2$. Here $R(x) = \sqrt{1+x^2}\, F(1/2, 1/2, 1; -x^2)$ and $F(...)$ is the hypergeometric function. Eq. (10) differs from Eq. (11) of Ref. [6] by the factor $R(\Omega/\omega_c)$. Notice that, if only one LL is occupied, $\tilde{k}_F \gg 1$ implies $\omega_c/\Omega \gg 1$; it follows that $R(\Omega/\omega_c) \approx (1 + \Omega^2/4\omega_c^2) \approx 1$. That is, we can approximate $R$ by 1 only for $\omega_c/\Omega \gg 1$. For only the lowest spin-polarized LL occupied, the pure exchange or correlation contribution to the total single-particle energy $E_{0,k_x,-1}$ is absent. That is, in the SHFA we have $E_{0,k_x,-1} = \epsilon_{0,k_x,-1}$; it is an exact result and independent of the value of $r_0$.

As is usual in the SHFA [9], [12], we treat the screened Coulomb interaction $V^s$, in Eq. (5), in the static limit. However, an essential difference in obtaining $V^s$ is that we take into account the spatial inhomogeneity of the 2DEG along the $y$ direction. We will calculate $V^s(q_x, q_y; q'_y)$ within the random phase approximation (RPA), i.e., we neglect the effect of many-body interactions in $V^s$. To this end let us consider the statically screened, by the 2DEG, potential $\varphi(x, y; x_0, y_0)$ of an electron charge at $(x_0, y_0)$, $e\delta(\vec{r} - \vec{r}_0)$. All charges are assumed, for definiteness, within the 2D plane. The 2D-Fourier transform $\varphi(q_x, q_y; x_0, y_0)$ of $\varphi(x, y; x_0, y_0)$ obeys the integral equation

$$\varphi(q_x, q_y; x_0, y_0) = \frac{2\pi e}{\varepsilon q}\{e^{-i\vec{q}\cdot\vec{r}_0} + \frac{e}{2\pi L} \sum_{n_\alpha, n_\beta=0}^{\infty} \sum_{k_{x\alpha}} F_{\beta\alpha}$$
$$\times \int_{-\infty}^{\infty} dq_{y1} \varphi(q_x, q_{y1}; x_0, y_0)(n_\alpha k_{x\alpha}|e^{iq_{y1}y}|n_\beta k_{x\beta})(n_\beta k_{x\beta}|e^{-iq_y y}|n_\alpha k_{x\alpha})\}, \quad (11)$$

where $\sigma_\alpha = \sigma_\beta = 1$ when only the lowest spin-polarized level is occupied. Here $k_{x\beta} = k_{x\alpha} - q_x$, $F_{\beta\alpha} = (f_\beta - f_\alpha)/(\epsilon_\beta - \epsilon_\alpha + i\hbar/\tau)$, and $\tau \to +\infty$ is an adiabaticity parameter.

Due to the spatial homogeneity of the system along the $x$- axis, we look for solutions of Eq. (11) in the form $\varphi(q_x, q_y; x_0, y_0) = \varphi(q_x, q_y; y_0) \exp(-iq_x x_0)$. Then $\varphi(q_x, q_y; y_0)$ obeys Eq. (11) if we change $\varphi(q_x, q_y; x_0, y_0)$ to $\varphi(q_x, q_y; y_0)$ and $\exp(-i\vec{q}\vec{r}_0)$ to $\exp(-iq_y y_0)$. Taking the Fourier transform with respect to $y_0$ of this equation for $\varphi(q_x, q_y; y_0)$, we obtain

$$\varphi(q_x, q_y; q'_y) = \frac{2\pi e}{\varepsilon q}\{2\pi\delta(q_y + q'_y) + \frac{e}{2\pi L} \sum_{n_\alpha, n_\beta=0}^{\infty} \sum_{k_{x\alpha}} F_{\beta\alpha}$$
$$\times \int_{-\infty}^{\infty} dq_{y1} \varphi(q_x, q_{y1}; q'_y)(n_\alpha k_{x\alpha}|e^{iq_{y1}y}|n_\beta k_{x\beta})(n_\beta k_{x\beta}|e^{-iq_y y}|n_\alpha k_{x\alpha})\}. \quad (12)$$

For flat LLs, i.e., for $\Omega \to 0$ and fixed width of the 2DEG, we can carry out the sum over $k_{x\alpha}$ exactly as well as the integral over $q_{y1}$ if we use Eq. (4). Then, if $\nu = 1$, Eq. (12) leads to

$$\varphi(q_x, q_y; q_y^{'}) = 4\pi^2 e\delta(q_y + q_y^{'})/\varepsilon[q + 2r_0\ell_0^{-1}\exp(-q^2\ell_0^2/2)\sum_{m=1}^{\infty}\frac{(q^2\ell_0^2/2)^m}{m \cdot m!}]. \tag{13}$$

Eq. (13) shows that for $r_0 \ll 1$ the screening in a wide channel ($\Omega \to 0$) is weak. However, if $r_0 \sim 1$ this bulk screening is rather essential for $q \sim \ell_0^{-1}$ as the screened potential becomes smaller than the bare one, $4\pi^2 e\delta(q_y + q_y^{'})/\varepsilon q$. Eq. (13), after integration over $q_y^{'}$, gives the RPA static dielectric function [12], [13] for flat LLs.

For a narrow channel we solve Eq. (12) by iteration. This results in a power series in the small parameter $r_0$. Writing

$$V^s(q_x, q_y; q_y^{'}) = e\varphi(q_x, q_y; q_y^{'}) = \sum_{j=0}^{\infty} V_j^s(q_x, q_y; q_y^{'}), \tag{14}$$

and using Eq. (12), we see that $V_0^s(q_x, q_y; q_y^{'})$ is given by Eq. (7); for $j \geq 1$ we have

$$V_j^s(q_x, q_y; q_y^{'}) = \frac{e^2}{\varepsilon q L} \sum_{n_\alpha, n_\beta=0}^{\infty} \sum_{k_{x\alpha}} F_{\beta\alpha} \int_{-\infty}^{\infty} dq_{y1} V_{j-1}^s(q_x, q_{y1}; q_y^{'})$$
$$\times (n_\alpha k_{x\alpha}|e^{iq_{y1}y}|n_\beta k_{x\beta})(n_\beta k_{x\beta}|e^{-iq_y y}|n_\alpha k_{x\alpha}). \tag{15}$$

Using Eq. (7) and Eq. (15) we obtain

$$V_1^s(q_x, q_y; q_y^{'}) = \frac{2\pi e^4}{\varepsilon^2 q\sqrt{q_x^2 + (q_y^{'})^2}} \int_{-\infty}^{\infty} dk_{x\alpha} \sum_{n_\alpha, n_\beta=0}^{\infty} F_{\beta\alpha}$$
$$\times (n_\alpha k_{x\alpha}|e^{-iq_y^{'}y}|n_\beta k_{x\beta})(n_\beta k_{x\beta}|e^{-iq_y y}|n_\alpha k_{x\alpha}). \tag{16}$$

Substituting $V^s(q_x, q_y; q_y^{'})$ given by Eq. (14) in Eq. (5) we obtain

$$\epsilon_{0,k_x,1}^{ec} = \sum_{j=0}^{\infty} \epsilon_{0,k_x,1}^{(j)ec}, \tag{17}$$

where $\epsilon_{0,k_x,1}^{(0)ec} \equiv \epsilon_{0,k_x,1}^{ex}$ is determined by Eqs. (8)-(10) and

$$\epsilon_{0,k_x,1}^{(1)ec} = -\frac{e^4}{4\pi^2 \varepsilon^2} \int_{-k_F}^{k_F} dk_x^{'} e^{-a^2 k_-^2 \tilde{\ell}^2/2} \int_{-\infty}^{\infty} dk_{x\alpha} \sum_{n_\alpha, n_\beta=0}^{\infty} F_{\beta\alpha} |M_{n_\alpha n_\beta}(k_{x\alpha}, k_x, k_x^{'})|^2. \tag{18}$$

Here

$$M_{n_\alpha n_\beta}(k_{x\alpha}, k_x, k_x^{'}) = \int_{-\infty}^{\infty} dq_y \frac{e^{-q_y^2 \tilde{\ell}^2/4}}{\sqrt{k_-^2 + q_y^2}} e^{iaq_y k_+ \tilde{\ell}^2/2} (n_\alpha k_{x\alpha}|e^{-iq_y y}|n_\beta k_{x\alpha} - k_-). \tag{19}$$

Because it is assumed that only the lowest LL is occupied, we can rewrite Eq. (18) as

$$\epsilon^{(1)ec}_{0,k_x,1} = \epsilon^{(1)ec}_I(k_x) + \epsilon^{(1)ec}_{II}(k_x), \quad (20)$$

where

$$\epsilon^{(1)ec}_I(k_x) = -\frac{e^4}{4\pi^2\varepsilon^2}\int_{-k_F}^{k_F}dk'_x e^{-a^2k_-^2\tilde{\ell}^2}\int_{-\infty}^{\infty}dk_{x\alpha}R_I(k_{x\alpha},k_-)M_I^2(k_{x\alpha},k_x,k'_x), \quad (21)$$

and

$$\epsilon^{(1)ec}_{II}(k_x) = \frac{e^4}{4\pi^2\varepsilon^2\hbar\tilde{\omega}}\int_{-k_F}^{k_F}dk'_x e^{-a^2k_-^2\tilde{\ell}^2}\int_{-\infty}^{\infty}dk_{x\alpha}\sum_{n_\alpha=1}^{\infty}\frac{f_{0,k_{x\alpha}-k_-}}{n_\alpha \cdot n_\alpha!}$$
$$\times [|M_{n_\alpha}^+(k_{x\alpha},k_x,k'_x)|^2 + |M_{n_\alpha}^-(k_{x\alpha},k_x,k'_x)|^2]. \quad (22)$$

Here we have $R_I(k_x,q_x) = (f_{0,k_x-q_x} - f_{0,k_x})/(\epsilon_{0,k_x-q_x} - \epsilon_{0,k_x})$,

$$M_I(k_{x\alpha},k_x,k'_x) = 2\int_0^{\infty}dq_y\frac{e^{-q_y^2\tilde{\ell}^2/2}}{\sqrt{k_-^2+q_y^2}}\cos[aq_y(k_{x\alpha}-k_x)\tilde{\ell}^2], \quad (23)$$

and

$$M_{n_\alpha}^{\pm}(k_{x\alpha},k_x,k'_x) = \int_{-\infty}^{\infty}dq_y\frac{e^{-q_y^2\tilde{\ell}^2/2}}{\sqrt{k_-^2+q_y^2}}[\frac{-iq_y \pm ak_-}{\sqrt{2}/\tilde{\ell}}]^{n_\alpha}\exp\{\frac{iaq_y\tilde{\ell}^2}{2}[k_+ \pm (2k_{x\alpha}-k_-)]\}. \quad (24)$$

In Eq. (22), assuming $\Omega/\omega_c \ll 1$ and $\tilde{k}_F \gg 1$, due to the well-satisfied condition $|k_-|/k_F \ll 1$, we have used the approximation $\epsilon_{n_\alpha,k_{x\alpha}} - \epsilon_{0,k_{x\alpha}-k_-} \approx \hbar\tilde{\omega}\,n_\alpha$. Because of the symmetry of the problem we have $\epsilon^{(1)ec}_{I,II}(k_x) \equiv \epsilon^{(1)ec}_{I,II}(-k_x)$. Thus, it is sufficient to consider only $k_x \geq 0$, i.e., the right half of the channel.

## III. STRONG CORRELATIONS INDUCED BY SCREENING AT THE EDGES

In Eq. (21) we can make the approximation, confirmed for assumed $\tilde{k}_F \gg 1$ as by analytical treatment so by numerical calculations, see below, $R_I(k_{x\alpha},k_-) \approx (-2\tilde{m}/\hbar^2) \cdot \delta(k_{x\alpha}^2 - k_F^2)$. It is also possible, for the assumed conditions, to take $a \approx 1$. We further use the strong magnetic field limit $r_0 \ll 1$ Then integrating over $k_{x\alpha}$ we obtain

$$\epsilon^{(1)ec}_I(k_x) = \frac{e^4 m^* \omega_c^2}{4\pi^2\hbar^2\varepsilon^2\Omega^2 k_F}\int_{-k_F}^{k_F}dk'_x e^{-k_-^2\ell_0^2}[M_I^2(k_F,k_x,k'_x) + M_I^2(-k_F,k_x,k'_x)]. \quad (25)$$

Notice that for fixed $k_F$, or width $W$ of the channel, $\epsilon_I^{(1)ec}(k_x)$ is determined only by screening at the edges of the channel. The latter depends practically only on the slope of the energy dispersion, i.e., on the group velocity $v_g$ of the edge states [10] which, in the strong $B$ limit, can be well approximated by $v_g^H = \hbar^{-1}[\partial \epsilon_{0,k_x,1}/\partial k_x]_{k_x=k_F}$. Finally, for fixed $W$ further analytical treatment, confirmed by numerical calculations, shows that $\epsilon_I^{(1)ec}(k_x) \propto (1/v_g^H)$. This behavior follows from the factor $R_I(k_{x\alpha}, k_-)$ in Eq. (21). Then, for fixed $k_F$ (or $W$) and a strong magnetic field, $r_0 \lesssim 1$, the self-consistent $v_g$, renormalized due to the many-body interactions, can be essentially different from $v_g^H$. We obtain $\epsilon_I^{(1)ec}(k_x) \propto (1/v_g)$ if the energy dispersion is smooth on the scale of $1/\ell_0$. As will be shown below, the most important correlations are related to the strong screening by the edge states as represented by Eq. (25). Hence, we must evaluate $\epsilon_I^{(1)ec}(k_x)$ using Eqs. (23) and (25). We first obtain approximate analytical expressions for $\epsilon_I^{(1)ec}(k_x)$ in the interior region (bulk) of the QW and at its edge, $\tilde{k}_x \approx \tilde{k}_F$.

In the inner region of the QW we may assume $\tilde{k}_F - \tilde{k}_x \gg 1$ in Eq. (25). Evaluating $M_I(...)$ from Eq. (23) we obtain [14]

$$M_I(\pm k_F, k_x, k_x') \approx 2 \int_0^\infty dq_y \frac{\cos[q_y(\pm k_F - k_x)\ell_0^2]}{\sqrt{k_-^2 + q_y^2}} = 2K_0(|\pm k_F - k_x| \cdot |k_-|\ell_0^2). \tag{26}$$

Then Eq. (25) gives

$$\epsilon_I^{(1)ec}(k_x) = (Ry^*/2\Delta \tilde{E}_{F\uparrow})[1 - k_x^2/k_F^2]^{-1}, \tag{27}$$

where $Ry^* = e^4 m^*/\hbar^2 \varepsilon^2$ is the effective Rydberg, $\Delta \tilde{E}_{F\uparrow} = \Delta E_{F\uparrow}/\hbar \tilde{\omega}$, and where we used the result [14]

$$\int_0^\infty dx e^{-x^2} K_0^2(|\pm \tilde{k}_F - \tilde{k}_x|x) \approx \int_0^\infty dx K_0^2(|\pm \tilde{k}_F - \tilde{k}_x|x) = \pi^2/4|\pm \tilde{k}_F - \tilde{k}_x|. \tag{28}$$

In the middle of the channel we have $k_x^2/k_F^2 \ll 1$ and Eq. (27) gives $\epsilon_I^{(1)ec} \approx Ry^*/2\Delta \tilde{E}_{F\uparrow}$, which is equal to $Ry^*$ if the Fermi level is at $\hbar \tilde{\omega}/2$ above the bottom of the LL. From Eq. (27) we see that in the inner region of the QW the correlations induced by screening near the channel edges are strong and they become considerably stronger as the edges of the QW are approached. In agreement with the discussion given below Eq. (25), from Eq. (27) for fixed $k_F$ (or $W$) and $k_x$ (or position within the channel) but variable $\Omega$, we have $v_g^H \propto \Omega^2$ and $\epsilon_I^{(1)ec}(k_x) \propto (1/v_g^H)$. At the right edge of the QW Eqs. (25) and (23) give

$$\epsilon_I^{(1)ec}(k_F) \approx \frac{Ry^*}{16\Delta \tilde{E}_{F\uparrow}} [1 + 4\sqrt{2/\pi}\tilde{k}_F] \approx \frac{Ry^*\tilde{k}_F}{2\sqrt{2\pi}\Delta \tilde{E}_{F\uparrow}}. \qquad (29)$$

From Eqs. (29) and (27) it follows that $\epsilon_I^{(1)ec}(k_F)/\epsilon_I^{(1)ec}(0) \approx \tilde{k}_F/\sqrt{2\pi}$. The first term in the brackets of Eq. (29) is related to screening at the left edge and is $4\sqrt{2/\pi}\tilde{k}_F$ times weaker than the second term describing screening at the right edge. Notice that the screening at the left edge is proportional to the rather weak suppressing factor $\ell_0/W$ in comparison with that at the right edge. It also demonstrates that screening by the edge states can cause strong correlations at distances, from the edge, much larger than $\ell_0$. Also, from Eq. (27) we see that in the middle of the parabolic channel, $k_x = 0$, for fixed $\Delta \tilde{E}_{F\uparrow}$ and $\Omega \to 0$ entailing $W \propto \Omega^{-1}$, we have a finite $\epsilon_I^{(1)ec}$ independent of $W$. In this limit though the distance from $y_0(k_x = 0) = 0$ to the edges, $W/2$, increases with $\Omega^{-1}$, $v_g^H$ decreases with $\Omega$ and their product gives a finite $\epsilon_I^{(1)ec}$.

The behavior of $\epsilon_I^{(1)ec}(k_x)$ as function of $\tilde{k}_x = k_x\tilde{\ell}$ obtained numerically from Eq. (21), without approximations, and from Eq. (25) is shown in Fig. 1 by curves 1 and 2, respectively. The near coincidence of the two curves for actual $\tilde{k}_x$ values shows that the approximation involved in obtaining Eq. (25) from Eq. (21) is a good one. For the conditions assumed in Figs. 1 - 6, $\tilde{k}_x$ practically coincides with the dimensionless oscillator center $Y_0 = k_x\tilde{\ell}\omega_c/\tilde{\omega}$; hence we will refer to $\tilde{k}_x$ as the oscillator center as well. Further, curve 3 in Fig. 1 shows $\epsilon_{0,k_x,1}^{ex}$ given by Eq. (9). Here we assume $\omega_c/\Omega = 30$, $\tilde{k}_F = 30$, and $r_0 \equiv Ry^*/(e^2/\varepsilon\ell_0) = 1$; all energies are given in units of $e^2/\varepsilon\ell_0$. For a GaAs QW, which is assumed for all figures, we have $\varepsilon = 12.5$, $m^* = 0.067m_0$, and $Ry^* \approx 11.7 meV$, which is close to typical values of $e^2/\varepsilon\tilde{\ell}$, see also below. Even for rather large $2\Delta \tilde{E}_{F\uparrow} = 1$, curve 1 demonstrates that the correlations related to screening at the edges of the channel are important everywhere in the channel. They become more important for smaller $\Delta \tilde{E}_{F\uparrow}$ but are most essential near the edges as shown by comparing curves 1 and 3.

Curve 3 in Fig. 1 demonstrates that the edge velocity, when only the exchange contribution $\epsilon^{ex}_{0,k_x,1}$, given by Eq. (9), is taken into account and correlation contributions are neglected, is positive and logarithmically divergent

$$v_g^{ex}(k_x \approx k_F) = \frac{e^2}{2\pi\hbar\varepsilon} e^{-\tilde{k}_0^2/4} K_0(\tilde{k}_0^2/4) \approx \frac{e^2}{2\pi\hbar\varepsilon} \ln[\frac{8}{\tilde{k}_0^2 + 4\pi^2\ell_0^2/L^2}]; \quad (30)$$

here $\tilde{k}_0 = \tilde{k}_x - \tilde{k}_F$ and the approximation sign holds for $\tilde{k}_0^2/4 \ll 1$. We have introduced the very small value $4\pi^2\ell_0^2/L^2$ to avoid the divergence for very small $|\tilde{k}_0|$. In this case a finite and relatively small $v_g$, for $\omega_c \gg \Omega$, is obtained without explicit appeal to many-body interactions, when $\Delta E_{F\uparrow}$ is independent of $\Omega$, as $v_g = v_g^H(k_F) = \hbar k_F/\tilde{m} \propto \Omega/\tilde{\omega}$, which is negligibly small compared to $v_g^{ex}$ given by Eq. (30). That is, any finite $v_g^H(k_F)$, obtained in the Hartree approximation, will lead to a divergent $v_g^{ex}(k_F)$, when only exchange is taken into account. Hence the exchange itself, treated in the HFA, strongly avoids the smooth, quasi-flat behavior of the energy dispersion near the channel edges. However, curves 1 and 2 in Fig. 1 demonstrate that the correlations lead to the logarithmically divergent negative contribution to the total edge velocity. Therefore, the correlations due to the screening by the edge states, restore the smooth behavior of the energy dispersion near the edges. We show in the following Sec. IV how this is possible. From Eqs. (22) and (24), for $|\tilde{k}_0| \gg 1$, we obtain

$$\epsilon^{(1)ec}_{II}(k_x) \approx 2\epsilon^{(1)ec}_{II}(k_F) \approx [\pi^2/6 - (\ln 2)^2] Ry^*/2, \quad (31)$$

where we used the identity $^{[14]}$ $\sum_{k=1}^{\infty} 1/(2^k k^2) = [\pi^2/6 - (\ln 2)^2]/2 \approx 0.58$. If we had considered only the term with $n_\alpha = 1$ in Eq. (22), then the RHS of Eq. (31) would be replaced by $Ry^*/2$; this shows that keeping only the term with $n_\alpha = 1$ is a good approximation. Considering only this term in Eqs. (22) and (24) we obtain

$$\epsilon^{(1)ec}_{II}(k_x) \approx Ry^*[1 + \Phi(\tilde{k}_F - \tilde{k}_x)]/4, \quad (32)$$

where $\Phi(x)$ is the probability integral. Curve 4 in Fig. 1 represents $\epsilon^{(1)ec}_{II}(k_x)$ as given by Eq. (32). The correlations here are related to the screening by the 2DEG within the channel, i.e., the strong screening by the edge states is excluded. The contribution to the total $v_g$ following from Eq. (32) is finite and negative and its role is minor in comparison with the two divergent contributions discussed above. Notice also that $\epsilon_{II}^{(1)ec}(k_x)$ in Eq. (32) is practically independent from $\Omega$ as well as from $v_g^H$.

## IV. STRONG SUPPRESSION OF EXCHANGE SPLITTING BY CORRELATIONS

### A. Restoration of smooth energy dispersion at the edges

For $2\Delta\tilde{E}_{F\uparrow} = 1$ and $\Omega/\omega_c \ll 1$, the ratio of $\epsilon_I^{(1)ec}(0)$, from Eq. (27), to $|\epsilon_{0,0,1}^{ex}|$, from Eq. (10), is $\sqrt{2/\pi}\, r_0$. Further, the above treatment shows that correlations induced by the screening can be so strong near the edges that in the *strong magnetic field limit* we must have not only $r_0 = Ry^*/e^2/\varepsilon\ell_0 \ll 1$ but also $10 \times r_0$ essentially less than 1. In this limit using Eqs. (9), (20), and (17) we obtain $\epsilon_{0,k_x,1}^{ec} \approx \epsilon_{0,k_x,1}^{ex} + \epsilon_{0,k_x,1}^{(1)ec} = \epsilon_{0,k_x,1}^{ex}[1 + \epsilon_{0,k_x,1}^{(1)ec}/\epsilon_{0,k_x,1}^{ex}]$. This can be rewritten as

$$\epsilon_{0,k_x,1}^{ec} \approx \epsilon_{0,k_x,1}^{ex}/[1 - \epsilon_{0,k_x,1}^{(1)ec}/\epsilon_{0,k_x,1}^{ex}], \tag{33}$$

since in this limit $\epsilon_{0,k_x,1}^{(1)ec}/|\epsilon_{0,k_x,1}^{ex}| \sim r_0 \ll 1$; the total energy is $E_{0,k_x,1} = \epsilon_{0,k_x,1} + \epsilon_{0,k_x,1}^{ec}$. Then the RHS of Eq. (33) well approximates the main contributions to $\epsilon_{0,k_x,1}^{ec}$ related with exchange and correlations. We further assume that the approximation given by Eq. (33) is also valid for $r_0 \sim 1$ and $\epsilon_{0,k_x,1}^{(1)ec}/|\epsilon_{0,k_x,1}^{ex}| \gtrsim 1$; this is an essential point for the MLDA that we now start to introduce. In addition, for the general case we assume that $\epsilon_{0,k_x,1}^{(1)ec}$ is given by

$$\epsilon_{0,k_x,1}^{(1)ec} = \epsilon_{II}^{(1)ec}(k_x) + (v_g^H/v_g)\epsilon_I^{(1)ec}(k_x), \tag{34}$$

where $\epsilon_{II}^{(1)ec}(k_x)$ is given by Eq. (32), $\epsilon_I^{(1)ec}(k_x)$ by Eq. (25), and the factor $(v_g^H/v_g)$ takes into account the real slope $v_g = \hbar^{-1}[\partial E_{0,k_x,1}/\partial k_x]_{k_x=k_F}$ of the total energy dispersion at $k_x = k_F$ (see the discussion following Eq. (25)). Eq. (34) can be well justified for strong $B$ and $r_0 \leq 1$ when the dependence of $\epsilon_{0,k_x,1}^{(1)ec}$ on $k_x$ is smooth on the scale of $1/\ell_0$. As discussed in Sec. III, for fixed $k_F$ $\epsilon_{II}^{(1)ec}(k_x)$ is practically independent of $v_g$ and $\epsilon_{I}^{(1)ec}(k_x)$ behaves as $1/v_g$. As for Eq. (33), its validity for $r_0 \sim 1$ is not obvious beforehand because both $\partial \epsilon_{0,k_x,1}^{(1)ec}/\partial k_x$ and $\partial \epsilon_{0,k_x,1}^{ex}/\partial k_x$ are logarithmically divergent at $k_x = k_F$. However, numerical calculations, cf. Figs. 2-5, show that for $r_0 \sim 1$ $\epsilon_{0,k_x,1}^{(1)ec}$, as given by Eq. (33), depends smoothly on $k_x$ even in the important edge region $k_x \approx k_F$.

In the *strong magnetic field limit* we have $v_g^H/v_g \approx 1$ if the confining potential is not too smooth. In the general case, for given $v_g^H$, $k_F$, $\omega_c$, etc., to calculate self-consistently the energy dispersion we should solve the equation

$$v_g = v_g^H + \frac{\partial}{\hbar \partial k_x} \{\epsilon_{0,k_x,1}^{ex}[1 - (\epsilon_{0,k_x,1}^{ex})^{-1}[\epsilon_{II}^{(1)ec}(k_x) + (v_g^H/v_g)\epsilon_{I}^{(1)ec}(k_x)]]^{-1}\}, \quad (35)$$

for $v_g$ and then substitute it in Eq. (4).

The approximation used in obtaining Eqs. (33)-(35) is based essentially on: (i) the assumption that the energy dispersion is smooth on the scale of $1/\ell_0$ as well as for $k_x \approx W/2\ell_0^2 \approx k_F$, (ii) the fact that $\epsilon_{I,II}^{(1)ec}(k_x)$ is practically independent of the changes of the eigenfunctions, when the smooth, on the scale of $\ell_0$, confining potential changes while $W$ is fixed, and, for such conditions, (iii) the strong dependence of $\epsilon_{I}^{(1)ec}(k_x) \propto (1/v_g)$ on $v_g$. This velocity can vary essentially as a result of changes in the confining potential which leave the electron density practically unchanged in the channel and even at the edges.

In line with the local-density approximation (LDA) [9] and [15], which includes exchange and correlation effects within a self-consistent framework, we assume that the energy dispersion relation given by Eqs. (33)-(35) can be obtained approximately by solving the single-particle Schrodinger equation (for $\sigma = 1$) with the Hamiltonian $\hat{h} = \hat{h}^0 + V_{XC}(y)$, where the self-consistent exchange-correlation potential is

$$V_{XC}(y) = \epsilon_{0,y/\ell_0^2,1}^{ec}, \quad (36)$$

and the function $\epsilon_{0,x,1}^{ec}$ is determined by Eqs. (33)-(35). Assuming that $V_{XC}(y)$ is smooth on the scale of $\ell_0$ we find, neglecting small corrections, that the corresponding energy dispersion is given again by Eqs. (33) - (35) for the lowest occupied spin-polarized LL. This confirms the self-consistency of our approximate study of the present many-body problem. However, in contrast with the LDA, our $V_{XC}(y_1)$ depends essentially on the slope $d[V_y + V_{XC}(y)]/dy$ at the edges $y \approx \pm y_0(k_F)$, which can be substantially away from $y_1$ as expressed by $|y_1 \pm y_0(k_F)|/\ell_0 \gg 1$. Because of that we refer for a strong magnetic field, $r_0 \lesssim 1$, to the approximation involved in Eqs. (33)-(36) as a modified LDA (MLDA). In contrast with the LDA, within the MLDA changes of the confining potential, which leave the electron density practically unchanged in the channel, can strongly modify the energy dispersion etc.

Though we consider only such the QW electron systems, that with correlations neglected are described by the Fermi level, when correlations are included, e.g., in the MLDA, the occupied ($n = 0$, $\sigma = 1$) LL and the empty ($n = 0$, $\sigma = -1$) LL can have different quasi-Fermi levels or the same quasi-Fermi levels which then coincides with the Fermi level. For $r_0 \sim 1$, typical in experiments, and referring to Fig. 1, we see that Eq. (33) gives a strong suppression of the exchange splitting between the ($n = 0, \sigma = 1$) and ($n = 0, \sigma = -1$) LLs. This is further shown in Figs. 2 and 3 for $\tilde{k}_F = 30$, $\omega_c/\Omega = 30$, and $2\Delta\tilde{E}_{F\uparrow} = 1$.

In Fig. 2 we plot $\epsilon^{ec}_{0,k_x,1}$, using Eq. (33), for $r_0 = 1$, $2/3$ and $3/2$ corresponding to curves 1, 2, and 3, respectively. Curve 4 represents the smooth parabolic dependence of $\epsilon_{0,k_x}$ given by Eq. (2), for $r_0 = 1$, shifted downward for an easier comparison with curve 1. It is clear that the correlations strongly suppress the sharp energy dispersion at the edges, caused by exchange, and restore a finite group velocity. Moreover, for $\tilde{k}_x \approx \tilde{k}_F$, curve 1 gives an even flatter behavior than that of curve 4.

In Fig. 3, for $r_0 = 1$ and the other parameters as in Figs. 1 and 2, we plot $E_{0,k_x,1} = \epsilon_{0,k_x,1} + \epsilon^{ec}_{0,k_x,1}$ (curve 1), $E_{0,k_x,-1} = \epsilon_{0,k_x,-1}$ (curve 2), and $E^{(0)}_{0,k_x,1} = \epsilon_{0,k_x,1} + \epsilon^{ex}_{0,k_x,1}$ (curve 3). Here $\epsilon^{ec}_{0,k_x,1}$ is given by Eq. (33), $\epsilon_{0,k_x,\pm1}$ by Eq. (2), and $\epsilon^{ex}_{0,k_x,1}$ by Eq. (9). We assume that the upper spin-split LL ( curve 2) is not occupied and the lower spin-split LL is occupied only for $\tilde{k}_F \leq 30$, when correlations are taken into account (curve 1) and when they are not (curve 3). The bare spin-splitting ($\propto g_0$) can practically be neglected due to its very small value in comparison with the gaps.

Comparing curves 1 and 3 in Fig. 3 shows that the correlations indeed change the in-finitely sharp energy dispersion at the edges into a smooth one such that the velocity $v_g$, for curve 1, is close to that without any many-body interaction, $v_g^H$, for this parabolic confinement, compare with the slope of curve 2 at $\tilde{k}_x = 30$. This is also expected from more detailed analytical considerations. Because of that we assumed $v_g^H/v_g \approx 1$ in Figs. 1 - 3 .

### B. Destruction of the $\nu = 1$ QHE state in a QW

Figure 3 shows that the correlations suppress very strongly the exchange-induced spin-splitting. Now if we neglect correlations and the $\nu = 1$ QHE state in the QW is possible, because the Fermi level (curve 5) is below the bottom of the $(n = 0, \sigma = -1)$ LL (curve 2), then with correlations taken into account the quasi-Fermi level of the $(n = 0, \sigma = 1)$ LL ( curve 4 ) is well above the bottom of curve 2; then the $\nu = 1$ QHE state is impossible. That is, the correlations lead not only to a strong suppression of the spin-splitting but also to a possibility of total destruction of the $\nu = 1$ QHE state, related to the spin-splitting of the $n = 0$ LL.

Now we apply our theory to the conditions of the experiments of Ref. [8] in GaAlAs/GaAs QWs for which $g_0 = -0.44$. The estimated QW parameters [8] for sample 1 are $W \approx 0.3\mu$m, $\hbar\Omega \approx 0.65 meV$, a linear density $n_L = n_S W \approx 7 \times 10^6 cm^{-1}$, where $n_S$ is the strong $B$ 2D electron density, and the $\nu = 1$ plateau structure is absent. For sample 2 the estimated parameters [8] are $W \approx 0.33\mu$m, $n_L B \approx 5 \times 10^6 cm^{-1}$, and $\hbar\Omega \approx 0.26 meV$. In this sample, with smaller $n_L$, $\hbar\Omega$, but almost the same $W$ as sample 1, the wide $\nu = 1$ plateau develops and is centered at $B = 7.3$T; this gives $\omega_c/\Omega \approx 45$ and $r_0 \approx 1.0$. The corresponding numbers for sample 1 are $B = 10.0$T, $\omega_c/\Omega \approx 25$, and $r_0 \approx 0.85$. Because the physical widths of both samples are close to each other we can assume $\tilde{k}_F \approx 15$ in both samples . This approximation has only a small quantitative effect on the results of a more precise treatment but simplifies their presentation considerably. In Figs. 4 and 5 we represent the energies in units of $\hbar\omega_c$ and measure them from the bottom of the $(n = 0, \sigma = -1)$ LL assumed empty. In Fig. 4 the parameters are those of sample 1 and in Fig. 5 those of sample 2.

In curve 1 of Fig. 4 we plot $E_{0,k_x,-1}/\hbar\omega_c = \tilde{k}_x^2/2(25)^2$ for the upper spin-split LL. Curve 2 shows $E_{0,k_x,1}$ obtained self-consistently from Eqs. (33)-(35), i.e., in the MLDA in which $v_g^H/v_g \approx 0.1$. Curve 3 is the quasi-Fermi level when only the states with $\tilde{k}_x \leq \tilde{k}_F = 15$ in curve 2 are occupied; notice that here in the MLDA we obtain the thermodynamically nonequilibrium state (though it is stationary in the MLDA) which, hence, can not be really observed. Also this thermodynamically nonequilibrium state (with different quasi-Fermi levels for the occupied LL and the other, empty, LL) do not have a finite gap that could lead to the $\nu = 1$ QHE. Curve 4 shows the same quantity as curve 2 but without the bare spin-splitting $g_0\mu_B B$. Curve 5 represents the lowest spin-split level, when the correlation interaction is omitted, i.e., $E_{0,k_x,1}^{(0)} = \epsilon_{0,k_x,1} + \epsilon_{0,k_x,1}^{ex}$. Finally, curve 6 shows the Fermi level, for $\tilde{k}_x \leq \tilde{k}_F = 15$, when correlations are neglected.

In curves 1-6 in Fig. 5 we plot the same quantities as in Fig. 4 but for the parameters of sample 2. In difference with Fig. 4, in the MLDA here we calculate a finite gap between the occupied states of the $(n = 0, \sigma = 1)$ LL and the empty $(n = 0, \sigma = -1)$ LL, given by curve 1, hence, here curve 3 is the corresponding Fermi level. For curve 2 we calculate self-consistently $v_g^H/v_g \approx 0.2$. Curve 7 gives the same behavior as curve 2 but without correlations related to the screening in the inner part of the 2DEG, $\epsilon_{II}^{(1)ec}(k_x)$. It is seen that $\epsilon_{II}^{(1)ec}(k_x)$ plays a minor role in the overall behavior of $E_{0,k_x,1}$ especially near the edges.

From Figs. 4 and 5 it follows that if the $\nu = 1$ QHE state, with only one LL occupied, cannot be achieved in sample 1 ( curve 3 intersects curve 1 in Fig. 4), then the $\nu = 1$ QHE state, with only one LL occupied, is possible in sample 2 because the Fermi level in Fig. 5 (curve 3) is slightly below, by $\Delta E_{\downarrow F} > 0$, the bottom of the $(n = 0, \sigma = -1)$ LL ( curve 1). The self-consistent results in Fig. 5 show that there is an activation gap $\Delta E_{\downarrow F} \approx 1.1 g_0 \mu_B B \approx 2K$, in nice agreement with that measured in Ref. [8] for $\nu = 1$. Notice that in Fig. 4 we have $\Delta E_{\downarrow F} < 0$ and $|\Delta E_{\downarrow F}| \approx 5.3 g_0 \mu_B B$; obviously $\Delta E_{\downarrow F} = -\Delta E_{F\downarrow} \leq 0$ corresponds to the absence of a finite gap pertinent to the $\nu = 1$ QHE state. Possible small flactuations of the bottom of the $(n = 0, \sigma = -1)$ LL, due to static random potentials in a real sample, will lead to a smaller effective $\Delta E_{\downarrow F}$. The latter should be compared with the value $E_R \approx 1K$ measured in Ref. [8]. We conclude that our theory explains well the observation, in Ref. [8], of a strong suppression of the spin-splitting in long quasi-ballistic GaAlAs/GaAs QWs.

Finally, in Fig. 6 we plot the effective, spatially inhomogeneous, g-factor $g^*_{op} = (E_{0,k_x,-1} - E_{0,k_x,1})/\mu_B B$ as a function of $\tilde{k}_x$. For convenience we take $g^*_{op}$ as positive. Curve 1 is obtained from curves 1 and 2 in Fig. 5 and curve 2 from curves 1 and 7; that is, curve 2 does not takes into account the relatively small correlations caused by the "bulk" screening of the 2DEG, $\epsilon^{(1)ec}_{II}(k_x)$. Curve 3 is obtained from curves 1 and 2 in Fig. 4 is given for comparison, though the pertinent state in Fig. 4 is not thermodynamically stable.

We call the g-factor $g^*_{op}(k_x) \equiv g^*_{op}(y_0(k_x))$ "optical" because it is related to the spin-splitting between states with the same $k_x$. In addition, due to the smooth dependence of $g^*_{op}(y_0)$ on $\ell_0$, we can approximate its spatial dependence by $g^*_{op}(y)$. From Figs. 4-6 it is seen that $g^*_{op}$ is essentially spatially inhomogeneous and near the channel edges it can be suppressed very strongly. Moreover, such an "optical" g-factor can be substantially different from the g-factor $g^*_{ac}$ deduced by the activated behavior of the conductance [8].

## V. CONCLUDING REMARKS

The above treatment was mainly devoted to QWs with $W \lesssim 0.3\mu$m. However, the main results can be directly extended to the regions close to the edges of substantially wider channels. This holds when the confining potential, without many-body interactions, can be approximated by $V'_y$. The treatment shows that for such channels the "optical", effective g-factor $g^*_{op}$, corresponding to a spin-splitting of the states with the same oscillator center $\tilde{k}_x$, is essentially spatially inhomogeneous in the range of many $\ell_0$ from the channel edge. It is also strongly suppressed in this region due to strong correlations. Such effect is of essential experimental interest [5] and appears very important when combined with the edge-state picture of the QHE [16] or the picture of the breakdown of the QHE developed in Ref. [10].

Though our calculations assume that the second spin-polarized ($\sigma = -1$) band is empty it does not mean that we must necessarily obtain only the state that can be described by the Fermi level. Our treatment can describe the stationary state with **different quasi-Fermi levels** for the occupied ($\sigma = 1$) and unoccupied ($\sigma = -1$) bands. This possibility is related to the properties of the SHFA which are similar to those of the HFA. In general, the ($\sigma = 1$) and ($\sigma = -1$) bands can have either a) different quasi-Fermi levels, when there is no finite gap between the occupied states of the ($\sigma = 1$) band and the bottom of the ($\sigma = -1$) one, or b) the same Fermi level, when there is a gap. Because of that we call curve 4, in Fig. 3, the *quasi-Fermi level* and curve 5 the Fermi level. In the first case $\sigma = -1$ LL has empty states below the quasi-Fermi level of the $\sigma = 1$ LL whereas in the latter the bottom of the empty LL is above the Fermi level.

We have treated only such electronic QWs that, without correlations, have the Fermi level below the bottom of the $\sigma = -1$ LL. This condition is valid for both sample 1 and sample 2 of Ref. [8]. But when exchange and correlations are taken into account both cases a) and b) are possible in the MLDA and pertain, respectively, to sample 1 and 2 of Ref. [8]. Because in a) the system is in a stationary state that cannot be stable and, in addition, a finite gap pertinent to the $\nu = 1$ QHE state is obviously absent, we conclude that the QW electron system cannot be in the $\nu = 1$ QHE state. The absence of the latter was observed for sample 1 in Ref. [8]. We emphasize that in our formal proof we always assume that $\sigma = -1$ LL is empty. As shown above, sometimes this is in contradiction with both the stability of the QW electron system and the presence of a finite gap which is pertinent to the $\nu = 1$ QHE state. The implication, when the contradiction is strong, is that the $\sigma = -1$ LL must be at least partially occupied for a thermodynamically stable state to exist, which in turns leads to the conclusion that the $\nu = 1$ QHE cannot be realized in such a system.

We have neglected the possible spatial inhomogeneity of the background dielectric constant $\epsilon$, along the $z$ direction. In GaAlAs/GaAs QWs such inhomogeneity is relatively small. Treating the spatially inhomogeneous screening, by the 2DEG of a QW, in the RPA for $T = 0$, we have neglected the effect of scattering on the screening. The latter should be more important in the screening of the edge states though for QWs with high mobility [8] it will not change our results essentially. We have also neglected the possible screening influence of the gates or other free charges outside the spacer layer; this seems a reasonable approximation for the experimental conditions of Ref. [8]. Notice that formally we considered very long channels appropriate to the samples of Ref. [8].

Though we have used simple analytical forms of the confining potential, obtained in the Hartree approximation as the sum of the bare confining potential and of the Hartree potential, they often approximate well the confining potential in real QWs and many of the above results hold for potentials of different form that are smooth on the scale of $\ell_0$. We have neglected possible changes in the confining potential due to many-body interactions, in its part given by the Hartree potential, i.e., changes induced by exchange and correlations. However, in the proposed MLDA we take into account an additional single-particle "confining" potential $V_{XC}(y)$, caused by exchange and correlations, which changes the energy dispersion essentially.

## ACKNOWLEDGMENTS

This work was supported by NSERC Grants Nos. ISETR106 and OGP0121756. In addition, O G B acknowledges partial support by the Ukrainian SFFI Grant No. 2.4/665.

FIGURES

FIG. 1. Many-body contributions to the single-particle energy of the lowest spin-polarized LL ($n = 0$, $\sigma = 1$), obtained in the strong magnetic field limit, as a function of the oscillator center $\tilde{k}_x \approx k_x \ell_0$. In all figures parabolic GaAs QWs are considered and the $n = 0$, $\sigma = -1$ LL is assumed empty. Curves 1 and 2 show the correlation contribution $\epsilon_I^{(1)ec}$, caused by screening at the edges of the wire (Eqs. (21) and (25)). Curve 3 shows the exchange contribution $\epsilon_{0,k_x,1}^{ex}$ (Eq. (9)) and curve 4 the correlation contribution, caused by screening within the "bulk" of the wire (Eq. (32)). The LL is occupied only for $\tilde{k}_x \leq \tilde{k}_F = 30$. For all curves we take $\omega_c/\Omega = 30$ and formally assume $r_0 = 1$. The Fermi level is above the bottom of the occupied LL at $\hbar\omega_c/2$.

FIG. 2. Single-particle energy $\epsilon_{0,k_x,1}^{ec}$, including many-body contributions in the MLDA (Eq. (33)), as a function of $\tilde{k}_x$. The parameters are the same as those in Fig. 1. Curves 1, 2, and 3 correspond to $r_0 = 1$, $2/3$ and $3/2$, respectively. Curve 4 shows the smooth parabolic dependence of $\epsilon_{0,k_x}$ (Eq. (2)) for $r_0 = 1$, shifted downward for clarity by $\hbar\tilde{\omega}/2$.

FIG. 3. Total single-particle energies $E_{0,k_x,1} = \epsilon_{0,k_x,1} + \epsilon_{0,k_x,1}^{ec}$ (curve 1) and $E_{0,k_x,-1} = \epsilon_{0,k_x,-1}$ (curve 2) as a function of $\tilde{k}_x$ for $r_0 = 1$ and the other parameters as in Figs. 1 and 2. Curve 4 is the quasi-Fermi level. Curve 3 shows $E_{0,k_x,1}$ when correlations are neglected, $E_{0,k_x,1}^{(0)} = \epsilon_{0,k_x,1} + \epsilon_{0,k_x,1}^{ex}$, and curve 5 is the corresponding Fermi level. Comparing curves 1 and 3 shows that the correlations smoothen the infinitely sharp energy dispersion at the edges such that $v_g$, for curve 1, is close to $v_g^H$ obtained without many-body interactions. This $v_g^H$ coinsides with the slope of curve 2 at $\tilde{k}_x = \tilde{k}_F = 30$.

FIG. 4. Energies as a function of $\tilde{k}_x$ for the parameters of sample 1 of Ref. 8 and $\nu = 1$. Only the $(n = 0, \sigma = 1)$ LL is occupied. Curve 1 shows $E_{0,k_x,-1} = \epsilon_{0,k_x,-1}$ and curve 2 $E_{0,k_x,1} = \epsilon_{0,k_x,1} + \epsilon^{ec}_{0,k_x,1}$ obtained in the MLDA (Eqs. (33) to (36)); curve 3 is the corresponding quasi-Fermi level. Curve 4 shows $E_{0,k_x,1} = \epsilon_{0,k_x,1} + \epsilon^{ec}_{0,k_x,1}$ when the bare spin-splitting $|g_0|\mu_B B \approx 0.015 \times \hbar\omega_c$, is neglected. Curve 5 shows $E^{(0)}_{0,k_x,1} = \epsilon_{0,k_x,1} + \epsilon^{ex}_{0,k_x,1}$, when correlations are neglected, and curve 6 gives the corresponding Fermi level. The parameters are $W \approx 0.30\mu m$, $\hbar\Omega \approx 0.65 meV$, $n_L = n_S W \approx 7 \times 10^6 cm^{-1}$, $B = 10.0T$, $\omega_c/\Omega \approx 25$, $r_0 \approx 0.85$, and $\tilde{k}_F \approx 15$.

FIG. 5. Same as in Fig. 4 with the parameters of sample 2 of Ref. [8] and $\nu = 1$. Here $W \approx 0.33\mu m$, $\hbar\Omega \approx 0.26 meV$, $n_L \approx 5 \times 10^6 cm^{-1}$, $B = 7.3T$, $\omega_c/\Omega \approx 45$, $r_0 \approx 1.0$, and $\tilde{k}_F \approx 15$. Curve 7 gives the same energy as curve 2 but without correlations related to the bulk screening. In contrast with Fig. 4, when exchange and correlations are taken into account a finite gap appears between the $\sigma = 1$ and $\sigma = -1$ LLs, pertinent to the $\nu = 1$ QHE state, and curve 3 is the Fermi level.

FIG. 6. Effective, spatially inhomogeneous, g-factor $g^*_{op} = (E_{0,k_x,-1} - E_{0,k_x,1})/\mu_B B$ as a function of $\tilde{k}_x$. Curve 1 is obtained from curves 1 and 2 in Fig. 5 and curve 2 from curves 1 and 7 in Fig. 5. Thus curve 2 neglects the small correlations caused by the bulk screening. Curve 3 is obtained from curves 1 and 2 in Fig. 4.

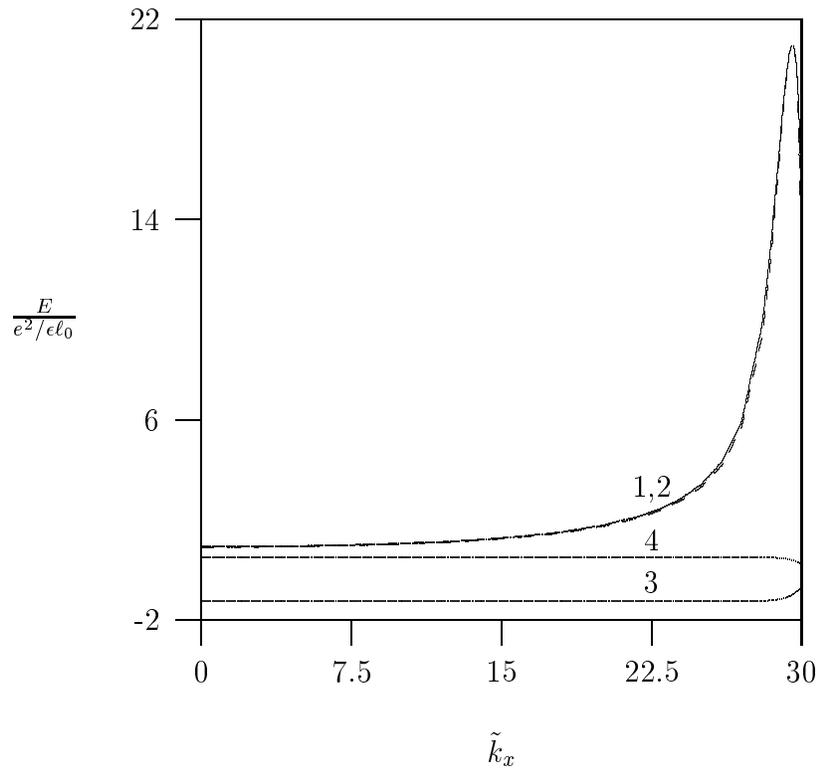

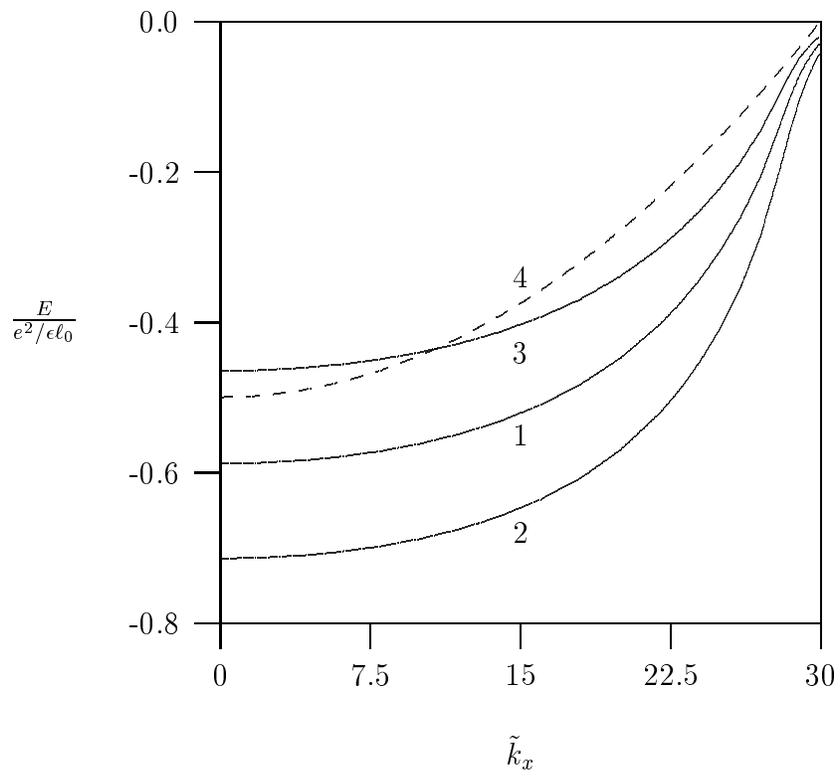

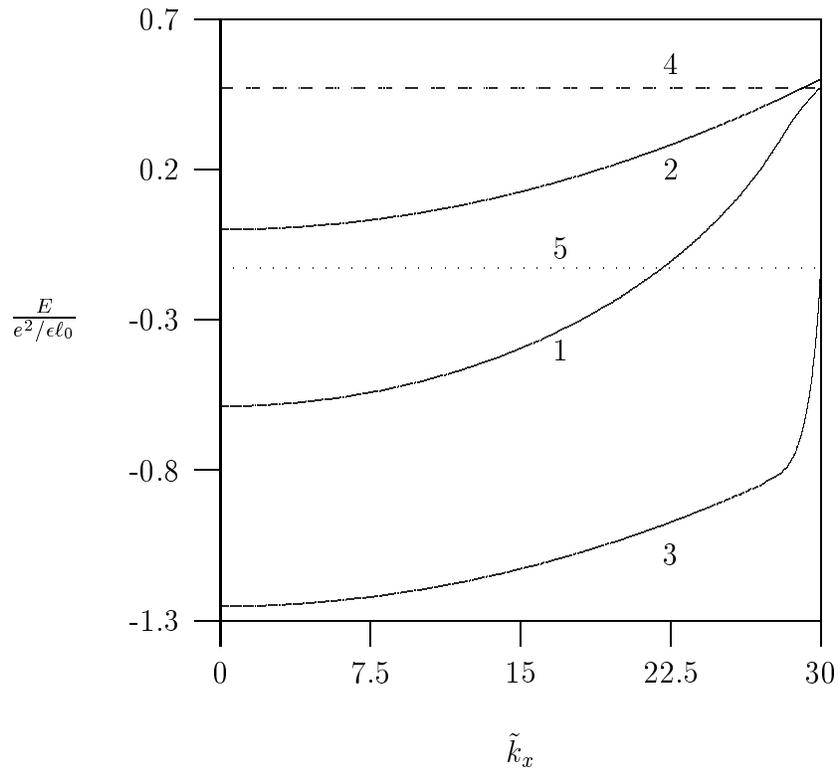

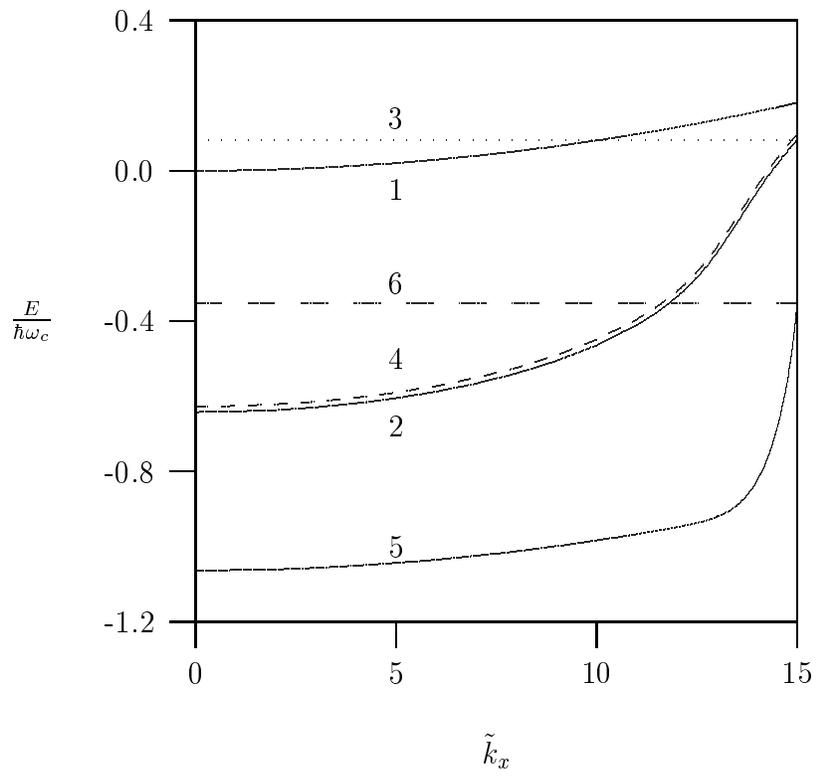

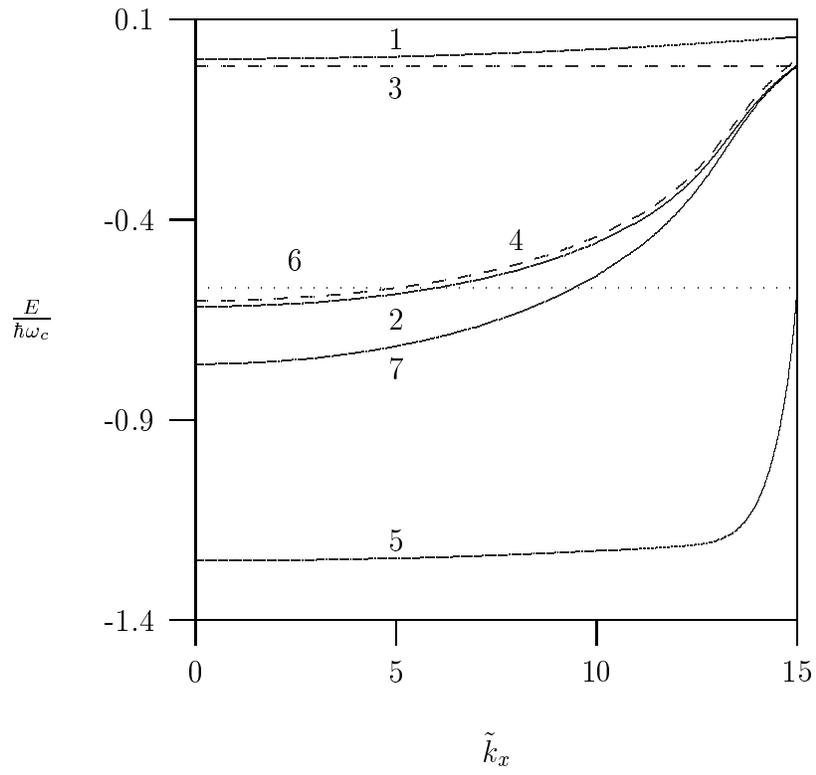

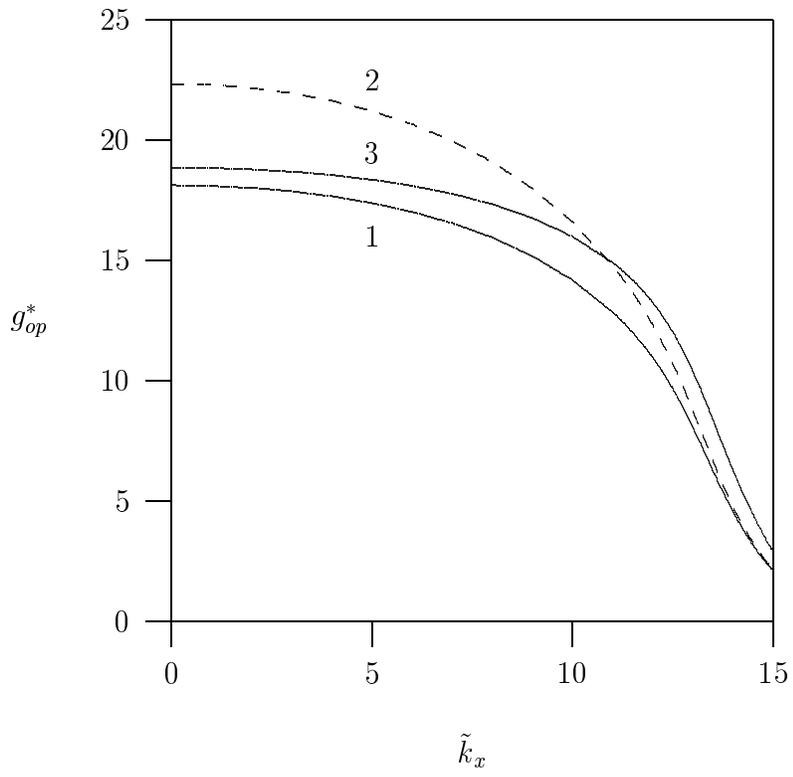